\documentclass[%
 aip,
 pop,%
 amsmath,amssymb,
 reprint,%
 numerical
]{revtex4-1}
\usepackage{graphicx}% Include figure files
\usepackage{bm}% bold math

 \usepackage{color}

\begin{document}
%opening
\title{Nonlinear evolution of the magnetized Kelvin-Helmholtz instability: from fluid to kinetic modeling}

\author{P.~Henri}
%\email[]{henri@df.unipi.it}
\altaffiliation{Dipartimento di Fisica ``E. Fermi'', Universit\`{a} di Pisa, Largo B. Pontecorvo 3, 56127 Pisa, Italy} 
\altaffiliation{Universit\'e de Nice Sophia Antipolis, CNRS, Observatoire de la C\^ote d'Azur, BP 4229 06304, Nice Cedex 4, France} 

\author{S.~S. Cerri} \altaffiliation{Dipartimento di Fisica ``E. Fermi'', Universit\`{a} di Pisa, Largo B. Pontecorvo 3, 56127 Pisa, Italy} \altaffiliation{Max-Planck-Institut f\"ur Plasmaphysik, EURATOM association, Boltzmannstr. 2, D-85748 Garching, Germany}

\author{F. Califano} \altaffiliation{Dipartimento di Fisica ``E. Fermi'', Universit\`{a} di Pisa, Largo B. Pontecorvo 3, 56127 Pisa, Italy}

\author{F. Pegoraro} \altaffiliation{Dipartimento di Fisica ``E. Fermi'', Universit\`{a} di Pisa, Largo B. Pontecorvo 3, 56127 Pisa, Italy}

\author{C.~Rossi} \altaffiliation{Dipartimento di Fisica ``E. Fermi'', Universit\`{a} di Pisa, Largo B. Pontecorvo 3, 56127 Pisa, Italy} \altaffiliation{LPP-CNRS, Ecole Polytechnique, UPMC, Universit\'e Paris VI, Universit\'e Paris XI, route de Saclay, 91128 Palaiseau, France}

\author{M.~Faganello} \altaffiliation{International Institute for Fusion Science/PIIM, UMR 7345 CNRS, Aix-Marseille University, 13397 Marseille, France}

\author{O. \v{S}ebek} \altaffiliation{Astronomical Institute and Institute of Atmospheric Physics, AS CR Bocni II/1401, CZ-14131 Prague, Czech Republic} \altaffiliation{Faculty of Nuclear Sciences and Physical Engineering, Czech Technical University in Prague, B\v{r}ehov\'{a} 7, 11519 Prague, Czech Republic}

\author{P. M. Tr\'{a}vn\'{i}\v{c}ek}  \altaffiliation{Space Sciences Laboratory, University of California Berkeley, 7 Gauss Way, Berkeley, CA 94720, USA} \altaffiliation{Astronomical Institute and Institute of Atmospheric Physics, AS CR Bocni II/1401, CZ-14131 Prague, Czech Republic}

\author{P. Hellinger} \altaffiliation{Astronomical Institute and Institute of Atmospheric Physics, AS CR Bocni II/1401, CZ-14131 Prague, Czech Republic}

\author{J. T. Frederiksen} \altaffiliation{Niels Bohr Institute, University of Copenhagen, Juliane Maries Vej 30, DK-2100 Copenhagen, Denmark}

\author{A. Nordlund} \altaffiliation{Niels Bohr Institute, University of Copenhagen, Juliane Maries Vej 30, DK-2100 Copenhagen, Denmark}

\author{S. Markidis} \altaffiliation{HPCViz Department, KTH Royal Institute of Technology, SE-100 44 Stockholm, Sweden}

\author{R. Keppens} \altaffiliation{Centre for mathematical Plasma Astrophysics, KU Leuven, Celestijnenlaan 200B, 3001 Leuven, Belgium}

\author{G. Lapenta} \altaffiliation{Centre for mathematical Plasma Astrophysics, KU Leuven, Celestijnenlaan 200B, 3001 Leuven, Belgium}

\date{\today}

\begin{abstract}
 The nonlinear evolution of collisionless plasmas is typically a multi-scale process where the energy is injected at large, fluid scales and dissipated at small, kinetic scales. Accurately modelling the global evolution requires to take into account the main micro-scale physical processes of interest. 
 This is why comparison of different plasma models is today an imperative task aiming at understanding cross-scale processes in plasmas. 
 We report here the first comparative study of the evolution of a magnetized shear flow, through a variety of different plasma models by using  magnetohydrodynamic, Hall-MHD, two-fluid, hybrid kinetic and full kinetic codes. 
 Kinetic relaxation effects are discussed to emphasize the need for kinetic equilibriums to study the dynamics of collisionless plasmas in non trivial configurations. 
 Discrepancies between models are studied both in the linear and in the nonlinear regime of the magnetized Kelvin-Helmholtz instability, to highlight the effects of small scale processes on the nonlinear evolution of collisionless plasmas. 
 We illustrate how the evolution of a magnetized shear flow depends on the relative orientation of the fluid vorticity with respect to the magnetic field direction  during the linear evolution when kinetic effects are taken into account. 
 Even if we found that small scale processes differ between the different models, we show that the feedback from small, kinetic scales to large, fluid scales is negligable in the nonlinear regime. This study show that the kinetic modeling validates the use of a fluid approach at large scales, which encourages the development and use of fluid codes to study the nonlinear evolution of magnetized fluid flows, even in the colisionless regime. 
\end{abstract}


\maketitle 

%######################
\section{Introduction}

In typical laboratory, space and astrophysical conditions, the nonlinear dynamics of magnetized plasmas is driven by the energy injected at large, fluid scales. The energy then cascades self-consistently towards smaller and smaller scales, until kinetic effects come into play. 
From a theoretical/modeling point of view, space plasmas represent a laboratory of excellence to study the physics of fundamental plasma processes, because of the wealth of in-situ diagnostics of improving quality accumulating in the form of electromagnetic profiles and particle distribution functions. The plasma turbulent state, routinely observed by satellites in the solar wind, is an archetype of plasma multi-scale behavior \cite{BrunoCarbone2005LRSP}. 

Many plasma processes naturally lead to a multi-scale dynamics, as for example at the interface between two different plasma regions, where large scale, fluid instabilities develop self-consistently and act as an energy source. 
This is the case for instance of the solar wind-magnetosphere interface, which plays a key role in the context of space weather modeling and forecasting. The connection between the solar wind and magnetosphere is mediated through the magnetosheath and magnetopause boundaries, that strongly depend on the solar wind properties and their variability. At the transition region between the solar wind flowing plasma and the Magnetosphere plasma at rest at low latitude, nearby the equatorial plane, the velocity shear between the two plasmas is an efficient source for the development of the Kelvin-Helmholtz instability (hereafter KHI). 

Satellite measurements have supplied clear evidence of rolled-up vortices at the flank of the Earth's Magnetopause \cite{Fairfield2000JGR,Hasegawa2004} where the KHI has been invoked to provide a mechanism by which the solar wind enters the Earth's magnetosphere, in particular to account for the increase of the plasma transport. Indeed, during northward magnetic field periods, when magnetic reconnection is considered as inefficient, at low latitude where the solar wind and magnetospheric fields nearly parallel, a relevant mixing between the solar wind and the magnetospheric plasma is observed, even larger than during southward configurations. KHI driven by a velocity shear can grow at low latitude since the magnetic field, nearly perpendicular to the plane where the instability develops,
does not inhibit its development. This provides an efficient mechanism for the formation of a mixing layer and for the entry of the solar plasma into the magnetosphere, explaining the efficient transport during northward solar wind periods. 
Other space observations have provided support for the development of the KHI in the environment of Saturn \cite{Mastersetal2010JGRA} and Mercury \cite{Sundbergetal2011PSS,Sundbergetal2012JGRA}. 

After saturation, the KH rolled-up vortices drive the formation of gradients at the ion inertial length and/or the ion Larmor radius up to electronic scales, and act as a source of secondary instabilities. 
In these conditions, previous studies have demonstrated that the nonlinear evolution of KH vortices enables the occurrence of magnetic reconnection driven by large-scale vortex motions, of interest for space plasmas\cite{FagCalPeg2008bPRL,FagCalPeg2008cPRL,FagPegCalMar2010PhPl,Henrietal2012POP}, as well as for astrophysical plasmas jets\cite{Keppens1999,Baty2003PhPl}.  
The vortex formation process indeed drags the magnetic field component parallel to the solar wind direction into the flow. As a result, the magnetic field is more and more stretched inside the vortices until it reconnects, redistributing the initial kinetic energy into accelerated particles and heating. 
Moreover, the density jump between the magnetosheath and magnetospheric plasmas drives fluid-like secondary instabilities such as the Rayleigh-Taylor instability \cite{Matsumoto2004GRL, FagCalPeg2008aPRL, Tenerani&al2011PPCF}. On top of that, the downstream increase of the magnetosheath velocity leads to super-magnetosonic regimes for which the KH vortices act as obstacles to the plasma flow, generating quasi perpendicular shocks \cite{Palermo&al2011JGRA,Palermo&al2011AnGeo,Henrietal2012POP}, thus modifying the transport properties of the plasma of the solar wind-magnetosphere. It is thus crucial to establish the role of these different secondary instabilities on the dynamics of the system, since they strongly influence the increase of the width of the mixing layer and its internal dynamics that are the most important factors for the evolution at the flank of the Earth's Magnetosphere. 

To summarize, the nonlinear evolution of the large scale fluid vortices is a fundamental plasma physics process driving the development of secondary instabilities. The energy then self-consistently cascades towards smaller and smaller scales, where the dynamics become kinetic, playing a significant role in the transport properties of the global system. 

Even if numerical studies of the nonlinear evolution of magnetized shear flows have been carried mainly by the means of fluid models (ideal/resistive MHD, Hall MHD, two-fluid), the increase of computational power has recently enabled to adress the problem of the kinetic modelling of shear flows in collisionless plasmas through hybrid PIC \cite{CoweeWinskeGary2009JGR,CoweeWinskeGary2010JGR} and full PIC simulations \cite{WilberWinglee1995JGR,Nakamuraetal2010POP,Nakamuraetal2011JGRA}. 
Low resolution simulations of the KHI have also been computed as a test problem to benchmark Vlasov codes \cite{Umedaetal2010PhPl}.  
% These works emphasize that
One of the main difficulty in the kinetic modeling of shear flows however remains the choice of the initial conditions. 
Indeed few kinetic equilibriums are known for shear flow configurations \cite{CaiStoreyNeubert1990PhFlB}.

To progress beyond the current state of the art in the nonlinear study of collisionless plasma evolution, the primary kinetic effects and physical processes at play should be understood through the complementary use of different models, from fluid to kinetic. 
At the end, only the kinetic modeling can validate or not the choice of a fluid approach. This is why a multi-model study is necessary to shade light on the fluid modeling of the nonlinear evolution of colisionless plasmas. 

We decided to focus in this work on the comparison of the evolution of a magnetized shear flow through the development of the KHI, using different plasma codes/models. However, we underline that the numerical modeling of plasma dynamics is a fundamental problem of major interest in present plasma physics research. Therefore, this study is of broad interest and is not limited to the KHI itself and related nonlinear dynamics. 

The paper is organized as follows. The different models and codes used in this study are described in section~\ref{section:model} and the configuration of the system under study is described in section~\ref{section:setup}. 
The results are presented in section~\ref{section:ComparativeStudy}: first, the kinetic relaxation to the initial fluid sheared flow equilibrium is discussed in section~\ref{section:kineticrelaxation}, then the linear growth rates of the magnetized Kelvin-Helmholtz instability are compared in section~\ref{section:linearphase}, the nonlinear phase of the magnetized Kelvin-Helmholtz is then studied for the different models in section~\ref{section:nonlinearphase}.
We finally discuss and conclude this work in section~\ref{section:conclusions}.

%##########################
\section{Models and codes}
\label{section:model}

The primary goal of this comparative study is to use different plasma models to solve a given physical problem, namely the evolution of a shear flow in a magnetized plasma. 
Multiscale properties of magnetized plasmas arise when nonlinear processes, hardly modeled by analytical studies, are at play. This is why numerical tools are needed to efficiently integrate in time the equations describing the plasma dynamics. 
On the other hand, benchmarking different numerical algorithms of the same model is out of the scope of this paper. 

%----------------------------------------------------------
\subsubsection{Hierarchy of plasma models}\label{section:models}
%----------------------------------------------------------

\begin{table}
 \begin{tabular}{ | l l | c | c | c | }
   \hline
   Model            &   & Closure           & Closure                 & Length                                    \\
                    &   & on ions           & on electrons            & scales                                    \\   
   \hline
   (i)   & MHD           & \multicolumn{2}{|c|}{Isothermal/Adiabatic}  & $L_{_{HD}}$                               \\
   (ii)  & Hall-MHD      & \multicolumn{2}{|c|}{           Adiabatic}  & $L_{_{HD}}$,$d_i$                         \\
   (iii) & Two-fluid     & Adiabatic         & Adiabatic               & $L_{_{HD}}$,$d_i$,$d_e$                   \\
   (iv)  & Hybrid        & \emph{no closure} & Isothermal              & $L_{_{HD}}$,$d_i$,$\rho_i$                \\
   (v)   & Full kinetic  & \emph{no closure} & \emph{no closure}       & $L_{_{HD}}$,$d_i$,$\rho_i$,$d_e$,$\rho_e$ \\
   \hline
  \end{tabular}
   \caption{Summary of the closure used in the different models for ions (first column) and electrons (second column). The length scales that are intrinsicaly part of the model are listed in the third column: $L_{_{HD}}$ the hydrodynamic scale of the system, $d_i$ and $d_e$ the ion and electron inertial lengths, $\rho_i$ and $\rho_e$ the ion and electron Larmor radii. }
   \label{table:closures}
\end{table}

The multiscale intrinsic nature of collisionless magnetized plasmas has lead to the development of a variety of plasma descriptions starting from a N-body description, through a mean field kinetic Vlasov description, to fluid descriptions (multi-fluid and magnetohydrodynamic). The underlying idea is that unresolved phenomena at small scales can be averaged out, leading to models much easier to handle than a N-body description. 

In this work, we have restricted our study to the following models: (i) magnetohydrodynamic (MHD), (ii) Hall-MHD, (iii) two-fluid, (iv) hybrid model with kinetic ions and massless electrons fluid, (v) full kinetic description of ions and electrons. 

The full kinetic model (full PIC code) contains all the relevant plasma length scales: the ion and electron inertial lengths $d_i$ and $d_e$, respectively, as well as the ion and electron Larmor radius $\rho_i$ and $\rho_e$, respectively. The hybrid model (hybrid PIC code) contains the ion length scales $d_i$ and $\rho_i$, while the two-fluid model contains the ion and electron inertial lengths $d_i$ and $d_e$. Ideal and resistive MHD are both transparent to the plasma lengths scales. This is summarized in table~\ref{table:closures}, third column. 

Concerning the closure of the models (summarized in table~\ref{table:closures}), the fluid codes (MHD, Hall-MHD and two-fluid) use either an isothermal or an adiabatic closure $\gamma = 1$ or $5/3$ respectively, the hybrid code uses an isothermal closure for electrons, while no closure is needed for the species described by a kinetic model: this is case for ions for the hybrid PIC code and for electrons and ions for the full PIC code. 

We stress that even if the kinetic models (full PIC in section~\ref{section:linearphase} and hybrid PIC in section~\ref{section:nonlinearphase}) should be considered the "models of reference", the initial equilibrium setup is chosen fluid-like, because of the difficulty of finding a Vlasov equilibrium, so that it does not describe a kinetic equilibrium for these two models. 
This is why the inter-model comparison should be done carefully, taking into account the fact that the force equilibrium can readjust during the initial phase.

The numerical algorithms used to solved the different plasma models are described in the next sections. (i) Two different codes are used to solve the MHD equations, 
they are described in sections~\ref{codes_stagger} and~\ref{codes_amrvac}, (ii) and (iii) the code solving the Hall-MHD and the two-fluid models is described in section~\ref{codes_twofluid}, (iv) the hybrid and (v) full kinetic models are solved using two different PIC algorithms described in sections~\ref{codes_hybrid} and~\ref{codes_ipic3d} respectively. 
These different codes have all been extensively tested and previously used to produce peer-reviewed scientific publications, so their validation is assumed and out of the scope of this paper.

%-------------------------------------------------
\subsubsection{MHD Stagger code}\label{codes_stagger}
%-------------------------------------------------

One of the MHD codes used in this work, the `Stagger Code', is a grid based, resistive and compressible MHD code. The code incorporates an adaptive hyper-resistivity and -viscosity scheme for enhanced control of dissipation introduced in fastmode waves, advective motion and shocks --- see for example \cite{BaumannGalsgaardNordlund2012} for an overview of the dissipation scheme and further code features. The MHD variables are defined on staggered grids, and the code conserves mass, momentum and $\nabla\cdot\mathbf{B}$ to machine precision. Interpolation of variables between different staggered grids is handled by using 5th order interpolation. Spatial derivatives are computed using 6th order accurate differential operators. The time integration of the MHD equations is performed using an explicit 3rd order low storage Runge-Kutta procedure \cite{1980JCoPh..35...48W}. For the MHD simulations conducted here, the resistive MHD equations solved in the Stagger Code are:

\begin{eqnarray}
\frac{\partial \rho}{\partial t}  &=& - \nabla \cdot (\rho \mathbf{u})  \label{equ:continuity} \\
\frac{\partial (\rho \mathbf{u})}{\partial t}  &=& - \nabla\cdot (\rho \mathbf{u} \mathbf{u} + \underline{\underline{{\bf \tau}}})
  - \nabla p \nonumber   + \mathbf{j} \times \mathbf{B} \label{equ:momentum} \\
\frac{\partial e}{\partial t}  &=& - \nabla \cdot (e \mathbf{u} + \mathbf{f_e})  - p \nabla \cdot \mathbf{u} \nonumber  \label{equ:energy} \\
\frac{\partial \mathbf{B}}{\partial t}  &=& - \nabla \times \mathbf{E}  \label{equ:induction} \\
\mathbf{j} &=& \nabla \times \mathbf{B}  \label{equ0:current} \\
\mathbf{E} &=& - \mathbf{u} \times \mathbf{B} + \eta \mathbf{j}  \label{equ:efield} \\
%Q_J &=& \eta j^2 \label{equ:Qmag}\\
p &=& (\gamma-1) e  \label{equ:pressure} \\
%\bar{\bar{\mathbf{\tau}}}_{ij} &=&
\tau_{ij} &=&  -\nu_{ij} \rho s_{ij} \label{equ:tauij} \\
s_{ij} &=&  \frac{1}{2}\left(\frac{\partial u_i}{\partial x_j} + \frac{\partial u_j}{\partial x_i} \right)   \label{equ:sij} \\
%Q_\nu &=& \rho\sum_{ij}\nu_{ij} s_{ij}^2
%  \label{equ:Qvisc}\\
\mathbf{f_e}   &=& - \nu_e \rho \mathbf{\nabla} (e/\rho)  \label{equ:fe}.
\end{eqnarray}

Here, $\rho$ is the mass density, $\mathbf{u}$ the bulk velocity, $p$ the pressure, $\mathbf{j}$ the current density, $\mathbf{B}$ the magnetic field, $e$ the thermal energy. $\nu_{ij}$ is the viscosity, $\nu_{e}$ is a thermal conductivity, and $\eta$ is the resistivity. $s_{ij}$ is the velocity shear tensor, 
%$\bar{\bar{\mathbf{\tau}}}_{ij}$ the viscous stress tensor. 
$\tau_{ij}$ the viscous stress tensor. 
$\mathbf{f_e}$ is a diffusive heat flux, needed for numerical stability. 
From Eqs.~(14) \&~(15) in Ref.~\cite{BaumannGalsgaardNordlund2012} using parameters $\nu1=0.007$, $\nu2=0.007$ and $\nu3=0.4$, the resistivity is $\eta\approx10^{-3}$ for both the adiabatic and isothermal cases with the setup described below. This initial resistivity will adjust during the instability growth and deviate from an approximate constant. This hyper-resistivity/-viscosity scheme will consequently yield larger effective values of the resistive and viscous terms, locally on the grid, when and where numerical critical structures appear. 
For the simulations with isothermal conditions, the energy equation is not active in the code and $\gamma=1$ (with $p=\rho$). The ideal gas law with $\gamma=5/3$ is assumed for the adiabatic case.
%-------------------------------------------------

%-------------------------------------------------
\subsubsection{MPI-AMRVAC}\label{codes_amrvac}
%-------------------------------------------------
The second code used in this work is the MHD module of the MPI-AMRVAC software~\cite{keppens2012}. The code uses a finite-volume discretization, combining explicit multi-step timestepping schemes with a variety of shock-capturing spatial discretizations. The code 
is designed to solve equations of generic form
\begin{equation}
\frac{\partial \mathbf{U}}{\partial t} + \nabla \cdot \mathbf{F(U) = S},
\end{equation}
and in this work, we employ two variants of the physical equations, which are (1) the isothermal, ideal MHD equations where conservative variables $\mathbf{U}=\left[\rho,\rho\mathbf{v},\mathbf{B}\right]$ include density, momentum and magnetic field; and (2) the full set of ideal MHD equations where in addition a total energy density equal to $\rho v^2/2 + B^2/2+p/(\gamma-1)$ is evolved. For the isothermal case, the pressure appearing in the momentum equation is at any time set from $p=0.5\rho$ in dimensionless units, such that it maintains the same equal uniform temperature conditions as initially in the full MHD case. The ratio of specific heats is set to $\gamma=5/3$ for the latter run. 
Although the code allows inclusion of physical sources \textbf{S}, such as viscosity or resistivity, in this work, we omit any explicitly added physical source 
that would introduce deviations from the pure conservation form. This means that any reconnection of the magnetic field is entirely due to the nonlinearities in the 
shock-capturing discretizations. To make the comparison possible with all other fixed grid simulations, we use MPI-AMRVAC in a domain decomposition approach, using only 1 grid of overall size $1536\times 512$, but employing 3072 grid blocks of size $16\times 16$ to exploit the MPI parallelism. The actual scheme used is an overall third-order Total Variation Diminishing (TVD) Lax-Friedrichs scheme, using a three-step Runge-Kutta variant combined with the third order \u{C}ada limiter~\cite{CadaTorrilhon2009JCP}. The $\mathbf{\nabla \cdot B}$ control is handled through a diffusive approach, introducing a (non-physical) source term diffusing monopole errors, which is handled in a split fashion. 
For the explicit time-advance, a courant parameter of 0.9 is used throughout. 
%-------------------------------------------------

%-------------------------------------------------
\subsubsection{Two-fluid code}\label{codes_twofluid}
%-------------------------------------------------
The Two-fluid code\cite{FagCalPeg2009NJP} is based on a two-fluid, ion-electron plasma approach including electron inertia effects in a fluid framework. 
The dimensionless equations of the model are obtained by using ion characteristic quantities: the ion mass $m_i$, the Alfv\'en  velocity  $c_A$ and  the ion  inertial scale length $d_i$ (in dimensionless units the electron inertial length reads $d_e^2 = m_e / m_i$). The density and motion equations read:
%%%%%%%%%%%%%%%%%%%%%%%%%%%
\begin{equation}
\frac{\partial n}{\partial t}+\nabla\cdot(n  \textbf{U})=0
\end{equation}
%%%%%%%%%%%%%%%%%%%%%%%%%%%
\begin{equation}
\frac{\partial (n \textbf{U})}{\partial t} +  \nabla \cdot \left[  n ( {\bf u}_i {\bf u}_i
+ d_e^2 {\bf u}_e {\bf  u}_e) + P_{tot} \underline{\underline{\ {\bf I}}} - {\bf BB} \right] = 0
\end{equation}
%%%%%%%%%%%%%%%%%%%%%%%%%%%
where ${\bf U} = {\bf u}_i + d_e^2 {\bf u}_e $ is the fluid velocity, $P_{tot}  = P_{th} + B^2/2$ is the total pressure, with $P_{th} = P_i + P_e$ the thermal pressure, and $n = n_i = n_e$ the density (we assumed quasi-neutrality). We adopt adiabatic closures for ions and electrons:
%%%%%%%%%%%%%%%%%%%%%%%%%%%
\begin{equation}
\frac{\partial ( n S_{e,i})}{\partial t}+\nabla \cdot(n S_{e,i} \textbf{u}_{e,i})=0
\end{equation}
%%%%%%%%%%%%%%%%%%%%%%%%%%%
where  $S_{e,i} = P_{e,i} n^{-\gamma}$. In the following we take $\gamma~=~5/3$.
The dimensionless sound velocity is defined as $c_s~=~\sqrt{\gamma P_{th} / n }$.
The electric field  is calculated by means of a generalized Ohm's law including electron inertia effects\cite{valentiniJCP2007}:
%%%%%%%%%%%%%%%%%%%%%%%%%%%%%%%%%
\begin{eqnarray*} 
 & (1 - d_e^2 & \nabla^2 ) \textbf{E} =  - \textbf{u}_e \times \textbf{B} - \frac{1}{n} \nabla  P_e    \\
 & & - d_e^2 \big[ \textbf{u}_i \times \textbf{B} - \frac{1}{n} \nabla P_i 
+ \frac{1}{n} \nabla\cdot [ n( \textbf{u}_i \textbf{u}_i - \textbf{u}_e \textbf{u}_e ) ] \big] \label{eq:ohm}
\end{eqnarray*}
%%%%%%%%%%%%%%%%%%%%%%%%%%%%%%%%%
The two-fluid code can also be run as a Hall-MHD code by imposing $d_e=0$.
Finally the magnetic  field  is calculated by solving the Faraday equation (Eq.~\ref{equ:induction})
%%%%%%%%%%%%%%%%%%%%%%%%%%%%%%%%%%%%%%%%%%%
%\[
%\frac{\partial \textbf{B}}{\partial t} = - \nabla \times \textbf{E}
%\]
%%%%%%%%%%%%%%%%%%%%%%%%%%%%%%%%%%%%%%%%%%%
and the current is given by ${\bf J} = n ({\bf u}_i- {\bf u}_e) = \nabla \times {\bf B}$ (we neglect the displacement current). 
These equations are integrated numerically in a 2D ($x, y$) slab geometry $-L_x/2 \leq x \leq L_x/2$, $0 \leq y \leq L_y$ using fully 3D fields (the so called 2.5D geometry). 
This code is based on a standard third-order Adams-Bashforth method for temporal discretization. It uses fast Fourier transform routines for spatial derivatives along the periodic $y$-direction and sixth-order compact finite difference scheme with spectral like resolution for spatial derivative along the inhomogeneous $x$-direction \citep{lele1992}. 
Numerical stability is achieved by means of filters, a spectral filter along the periodic $y$-direction and a sixth-order spectral-like filtering scheme along the inhomogeneous $x$-direction \citep{lele1992}. 
%-------------------------------------------------

%-------------------------------------------------
\subsubsection{Hybrid PIC code}\label{codes_hybrid}
%-------------------------------------------------
The hybrid code is based on a current advance method and a cyclic leapfrog algorithm \cite{Hybrid}. In the model, ions are treated by using a particle in cell scheme while electrons are represented by a massless, isothermal, charge neutralizing fluid. The code self-consistently solves equations of motion for ions 
\begin{eqnarray}
\frac{d\textbf{x}_{s}}{dt} & = & \textbf{v}_{s},\label{eqn-hyb-dxdt}\\
\frac{d\textbf{v}_{s}}{dt} & = & \frac{q_{s}}{m_{s}} \left(\textbf{E} + \textbf{v}_{s} \times \textbf{B}\right), \label{eqn-hyb-dvdt}
\end{eqnarray}
together with Faraday's law for magnetic field (Eq.~\ref{equ:induction}) and a generalized Ohm's law for electric field in the form
	\begin{eqnarray*} \label{eqn_hybrid_ohm}
	 \textbf{E} & = & \frac{1}{\rho_{c}} \left(\frac{(\nabla \times \textbf{B}) \times \textbf{B}}{\mu _{0}} -
	 \textbf{J}_{i}\times\textbf{B} - \nabla p_{e} \right) + \eta \left( \nabla \times \textbf{B}\right),
	\end{eqnarray*}
here $\textbf{x}_{s}$, $\textbf{v}_{s}$ are positions and velocities of particles of species $s$, $\rho_{c}=\sum\limits_s \rho_{c,s}$ and $\textbf{J}_{i} = \sum\limits_s \textbf{J}_{i,s}$ are total ionic charge and current densities, $\textbf{B}$ is the magnetic field, $\mu_{0}$ is vacuum permeability and $\eta$ is a resistivity parameter. The pressure of electrons is obtained as $p_{e}=\rho_e k_{B} T_e$ where the electron temperature $T_e$ is set as initial condition and remains constant during the simulation (the electron fluid is isothermal). The electron density is computed from the total ionic charge density using the assumption of quasineutrality $\rho_e=\rho_c/e$, here $e$ stands for elementary charge.
%-------------------------------------------------

%-------------------------------------------------
\subsubsection{Full PIC code}\label{codes_ipic3d}
%-------------------------------------------------
iPIC3D is a fully kinetic, fully electromagnetic Particle-in-Cell code \citep{iPIC3D}. It implements the moment implicit method \citep{brackbill1982implicit}. In this code, the second order formulation of Maxwell's equations for the electric field is discretized implicitly in time:
\begin{subequations}
 \label{Maxwell4}
 \begin{equation}
 \label{Maxwell4a}
 \mathbf{E^{n+1}} - (c  \Delta t)^2 \nabla^2  \mathbf{E^{n+1}} =
 \end{equation}
 \begin{equation*}
  \mathbf{E^{n}}  + c \Delta t  \Big[\nabla  \times \mathbf{B}^{n}
 - 4 \pi (\mathbf{J}^{n+1/2}  + c  \Delta t \nabla \rho^{n+1/2}  )\Big] .
 \end{equation*}
Once the electric field $ \mathbf{E^{n+1}}$ is calculated, the magnetic field is advanced in time solving the discretized Faraday's law:
\begin{equation}\label{Maxwell4b}
\frac{ \mathbf{B}^{n+{1}} - \mathbf{B}^{n} }{\Delta t}= - \nabla \times \mathbf{E}^{n+1} .
\end{equation}
\end{subequations}
The Maxwell's equations are differenced in space on a uniform cartesian grid and the simple box scheme is used for the spatial differentiation of spatial operators in the field equations (Eqs.(\ref{Maxwell4a})-(\ref{Maxwell4b})). The linear system arising from Eqs.(\ref{Maxwell4a})-(\ref{Maxwell4b}) is solved using the Generalized Minimal RESidual (GMRes) solver.
The equations of motion of particles are differenced in time using the implicit midpoint integration rule:
\begin{subequations} 
\label{eom}
\begin{eqnarray}
\mathbf{x}_p^{n+1} & = & \mathbf{x}_p^{n} +  \mathbf{v}_p^{n+1/2}\Delta t \\  
\mathbf{v}_p^{n+1} & = & \mathbf{v}_p^{n} + \frac{q_s}{m_s}(\mathbf{E}^{n+1}_{1/2} + \frac{\mathbf{v}_p^{n+1/2}\times \mathbf{B}^{n+1}_{1/2} }{c}) \Delta t
\end{eqnarray} 
\end{subequations}
$\mathbf{E}^{n+1}_{1/2}$, and $\mathbf{B}^{n+1}_{1/2}$ are the electric and magnetic field, calculated at the midpoint of the orbit and $\mathbf{v}_p^{n+1/2}$ is the average of $\mathbf{v}^{n}_p$ and $\mathbf{v}^{n+1}_p$. The iPIC3D code solves the particle equation of motion (Eq.(\ref{eom})) by an iterative method based on a fixed number of predictor-corrector iterations. 
%#########################################################################

%#########################################################################
\section{Description of the simulations setup} \label{section:setup}
%#########################################################################

The different models and corresponding codes previously described are used to integrate the linear and nonlinear evolution of a magnetized shear flow plasma unstable to the KHI (section~\ref{section:linearphase} and~\ref{section:nonlinearphase} respectively).

\begin{figure}[!t]
\begin{center}
 \includegraphics[width=\linewidth]{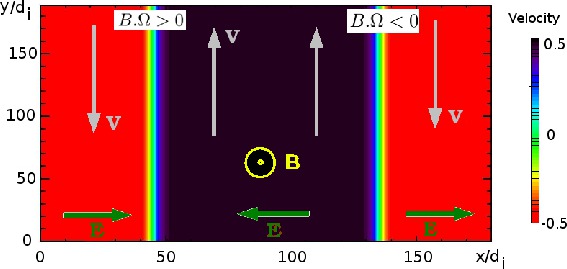}
\end{center} 
\caption{Initial double periodic setup for the Kelvin-Helmholtz benchmark. The direction of the velocity, magnetic and electric fields are shown. The magnetic field has also a small component in the $y$-direction (not shown here). Note that the layers are characterized by either $\mathbf{B} \cdot \mathbf{\Omega} > 0$ or $\mathbf{B} \cdot \mathbf{\Omega} < 0$. }
\label{fig:SetupKH}
\end{figure}

The simulation setup is identical for the different models. We consider a 2D $(x,y)$ physical space, with 3D vector fields and an initial MHD equilibrium (note: not a Vlasov equilibrium). In order to avoid additional/spurious effects due to different implementations of boundary conditions, a double shear layer is considered in order to impose periodic boundary conditions. The initial flow configuration is shown in Fig.~\ref{fig:SetupKH}. 

In the following, all quantities are normalized to ion quantities: the ion gyro-frequency, $\omega_{ci}$, the ion inertial length, $d_i$, and the Alfv\'en velocity $V_A$. The size of the numerical box is given by $L_x=180$ and $L_y=30 \ \pi$. The number of grid points in the XY-plane of the simulation is $N_x=1536$ and $N_y=512$. We use 1024 particles per cell in the hybrid PIC code and 200 particles per cell in the full PIC code. 

The initial velocity field $\mathbf U = U_y(x) \ \mathbf e_y$ contains a periodic double shear layer where the velocity varies from $-A_{eq}$ to $+A_{eq}$. The shear layers are located in $x_{c,1} = L_x / 4$ and $x_{c,2} = 3/4 \ L_x$. The velocity profile reads: 
\[ 
U_y(x) = A_{eq} \ \left[ \tanh \left( \frac{x - x_{c,1}}{ L_{eq} } \right) - \tanh \left( \frac{x - x_{c,2}}{ L_{eq} } \right) -1 \right]  ; 
\]
where the maximum velocity field strength is $A_{eq} = 0.5$ and the shear length scale is given by $L_{eq} = 3$ (in $d_i$ units), 
which implies $L_{eq} = 24 d_e$ in terms of the electron inertial length $d_e$ for the two-fluid and full-PIC simulations, using a mass ratio $m_p/m_e = 64$. 
The initial current is taken to be zero, $J_{eq} = 0$. The initial magnetic field is $\mathbf B_{eq} = B_0 sin(\theta) \ \mathbf e_y + B_0 cos(\theta) \ \mathbf e_z$, where $B_0 = 1$ and $\theta = 0.05$, so that $B_z = 20 \ B_y$. 

Note that the (mostly out-of-plane) magnetic field is constant, always pointing in the same direction. On the contrary, the direction of convective electric field $\mathbf E = - \mathbf U \times \mathbf B_{eq}$ varies from one layer to the other: it is directed towards (resp. away from) the shear layer at $x_{c,1}$ (resp $x_{c,2}$). Note also that the vorticity is parallel to the out-of-plane magnetic field ($\mathbf B_{eq} \cdot \mathbf \Omega > 0$, where $\mathbf \Omega = \mathbf \nabla \times \mathbf U$) at $x_{c,1}$, while it is anti-parallel ($\mathbf B_{eq} \cdot \mathbf \Omega < 0$) at $x_{c,2}$. 

The initial electron and ion pressures are isotropic, with $P_e = P_i = 0.5$, corresponding to a total thermal pressure $P_{tot} = P_e + P_i = 1$. The initial density is constant in the simulation box, $n = 1.0$, so that the electron and proton temperatures are equal, $T_e = T_i = 0.5$. 
Quasineutrality $n_i = n_e = n$ is imposed at the beginning of the simulation for the full PIC code, while it is assumed in the other models. Finally, in the two-fluid and full kinetic models, the proton-to-electron mass ratio is $m_p/m_e = 64$. 

%================================
%================================

An initial incompressible perturbation $\delta \mathbf U$ on the velocity field is imposed in the fluid models (not needed in the hybrid and full PIC codes) as follows:  
\[ \delta \mathbf{U} = \mathbf e_z \times \mathbf \nabla \psi \]
with 
 \[ \psi = \epsilon f(x) \sum_{m=1}^{N_y/4} cos(2 \pi m y / L_y + \phi_m ) / m \ ; \]
where
\[ f(x) = \exp \left[- ((x-x_{c,1}) / L_{eq})^2 \right] + \exp \left[ - ((x-x_{c,2}) / L_{eq})^2 \right] \]
%; xc = Lx / 2
and $\epsilon$ is such that $max(| \delta U |) \simeq 10^{-3}$ and $\phi_m$ are random phases. 

When studying the linear phase of the KHI (section~\ref{section:linearphase}), random phases $\phi_m$ are considered that may vary from one simulation to the other, since this does not modify the evaluation of the linear growth rate. 
On the contrary, identical random phases are considered for all runs, but different from one layer to the other, when studying the nonlinear phase of the instability (section~\ref{section:nonlinearphase}), since the initial phases influences the nonlinear evolution of the full system. 

The plasma beta is $\beta = 2$, so that the ion inertial length and the ion gyroradius are equal. The setup has been chosen so that the initial shear length is close to the ion inertial length and the ion gyroradius, in order to study the transition between the fluid and kinetic regimes. The grid length has been chosen so that the fastest growing mode is m=2, this way two rolled-up vortices form in the simulation box. 
%#########################################################################

%##########################
\section{Comparative study} \label{section:ComparativeStudy}

The kinetic relaxation oberved with kinetic models is first studied in section~\ref{section:kineticrelaxation}. 
We then discuss the linear part of the KHI in section~\ref{section:linearphase} and its nonlinear part in section~\ref{section:nonlinearphase}. 
%================================
\begin{figure}[b!]
\begin{center}
 \includegraphics[width=\linewidth]{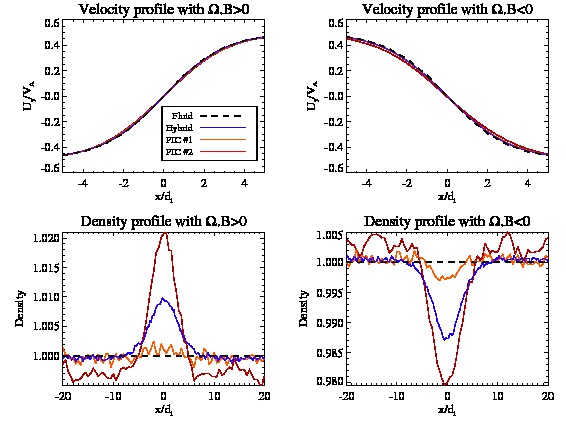}
 \end{center} 
\caption{Initial relaxation of the kinetic simulations due to the absence of a kinetic equilibrium. The fluid profile (MHD and two-fluid models) is shown with a black dashed line. The hybrid PIC and full PIC profiles are respectively shown by blue and red lines. 
To illustrate the influence of the plasma beta on this initial kinetic relaxation process, two full PIC simulations have been used: PIC $\#$1 and PIC $\#$2 simulations respectively use $U_{th,i}/V_A = 1/3$ and 1, so that $\beta_i = 0.1$ and 1 respectively, where $U_{th,i}$ is the ion thermal velocity.}
\label{fig:KinSetup}
\end{figure} 
%================================

%----------------------------------------
\subsection{Initial kinetic relaxation} \label{section:kineticrelaxation}
%----------------------------------------
In both kinetic simulations (hybrid and full PIC), the initial profiles are (asymmetrically) modified at the velocity shear location on the ion gyration time scale, before the KHI develops. This is due to the kinetic relaxation in response to the initial fluid (not Vlasov) equilibrium \cite{CaiStoreyNeubert1990PhFlB}. 
This phenomenon is particularly evident in the case $\mathbf{B} \cdot \mathbf{\Omega} < 0$ and becomes stronger as the plasma beta increases. This effect, illustrated by the average mean (fluid) velocity shear computed for times $5 < t \omega_{ci} < 60$ (Fig.~\ref{fig:KinSetup}, top panels), may explain part of the differences between the fluid and kinetic linear growth rates, since the fastly modified initial shear profile will respond differently to the imposed fluid profile. 

%================================
\begin{figure}[b!]
\begin{center}
 \includegraphics[width=\linewidth]{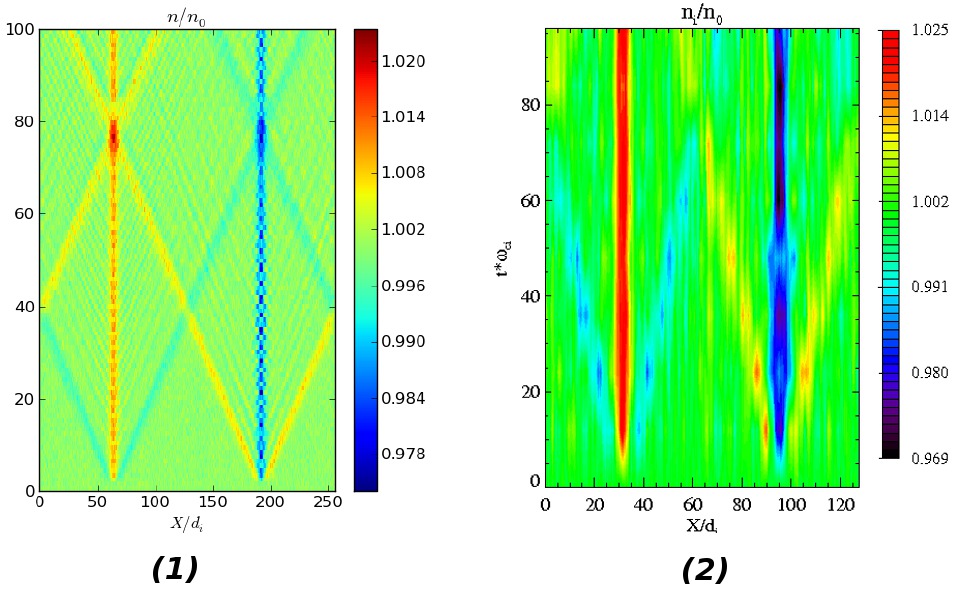}
 \end{center} 
\caption{Initial relaxation of the mean ion density profile in the kinetic simulations due to the absence of a kinetic equilibrium. Left panel: hybrid PIC simulation. Right panel: low resolution full PIC simulation (in a smaller box). The relaxation process is similar for both simulations. }
\label{fig:KinSetupEvolution}
\end{figure} 
%================================

Furthermore, Fig.~\ref{fig:KinSetup} also shows that the shear layer is modified more significantly in the full kinetic simulation than in the hybrid simulation. To quantify this feature, the layer with $\mathbf{\Omega} \cdot \mathbf{B} > 0$ broadens from the initial (fluid) thickness $L_{fluid}=3$ to $L_{Hyb}=3.06$ and $L_{Full}=3.16$ in the hybrid and the full ($\#$2) kinetic simulations respectively. While the layer with $\mathbf{\Omega} \cdot \mathbf{B} < 0$ broadens to $L_{Hyb}=3.14$ and $L_{Full}=3.45$ in hybrid and full kinetic simulations respectively. Note that both simulations are computed for the same ion beta $\beta_i = 1$. Here $L$ is computed as the distance from the layer center ($x_{c,1}$ or $x_{c,2}$) at which the velocity field reaches the value $U_{y}=\pm A_{eq}\tanh(1)$.  

This effect on the mean velocity goes together with a modification of the plasma density nearby the center of the velocity shear layer, due to finite Larmor radius effects. A `bump' for $\mathbf{\Omega} \cdot \mathbf{B} > 0$ (resp. a `dip' for $\mathbf{\Omega} \cdot \mathbf{B} < 0$) in the plasma density is generated in both the hybrid and the full PIC simulations, as shown in Fig.~\ref{fig:KinSetup}, bottom panels, where the blue and red lines respectively show the hybrid and the full PIC simulations and the black dashed lines show the fluid (MHD and two-fluid) profile.  

Since the initial condition is not a Vlasov equilibrium, a disturbance on the plasma density is excited and travels at constant speed through the simulation domain along the $x$-direction, as shown by two kinetic simulations (hybrid PIC and full PIC) in Fig~\ref{fig:KinSetupEvolution}. This localized travelling perturbation, of magnetosonic nature, excited at a given shear layer eventually reaches the other layer and may modify the dynamics of the system. This artificial effect is further increased by the double periodic boundary conditions that do not allow to get rid of these density perturbations. We do not observe in our hybrid and full simulations a clear relaxation of this disturbancy.

%================================
\begin{figure}[t!]
\begin{center}
    \includegraphics[width=\linewidth]{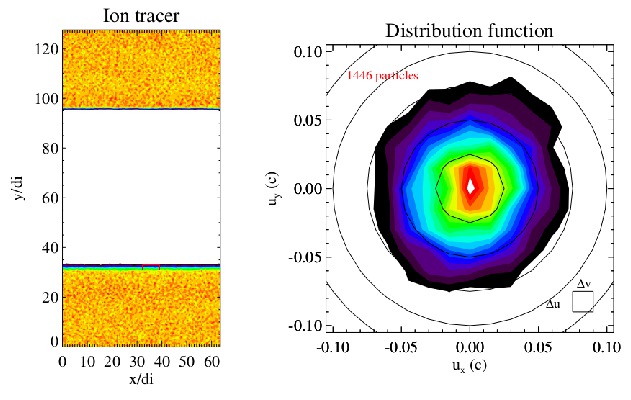}
 \end{center} 
\caption{Snapshot of the ion tracer at time $t \omega_{c,i} = 48$ (before the beginning of the Kelvin-Helmholtz instability) (left panel) for a PIC simulation with no in-plane magnetic field. The ion distribution functions (right panel) is computed at the bottom shear layers (inside the red box in the left panel). The black lines represent the expected gyrotropic contours of the distribution function. Note the generation of the temperature anisotropy $T_x < T_y$ corresponding to the layer $\mathbf{\Omega} \cdot \mathbf{B} < 0$. }
\label{fig:DistFunction}
\end{figure} 
%================================

Still considering the growing phase of the instability, we observe an initial deformation of the distribution functions in both the full and hybrid PIC simulations, which is interpreted in terms of the kinetic relaxation from an initial setup that is an MHD equilibrium but not a kinetic equilibrium. At the velocity shear layers location, the initially gyrotropic plasma becomes agyrotropic after a few gyro-periods (so that temperatures in $x$ and $y$ directions are different), as illustrated in Fig.~\ref{fig:DistFunction} where we show the result of the full PIC simulation. Note in the right panel the generation of a temperature anisotropy: $T_x < T_y$ corresponding to the bottom layer $\mathbf{\Omega} \cdot \mathbf{B} < 0$. The agyrotropy leads to the modification of the full pressure 
tensor\cite{Cerri2013} which is properly included only in the kinetic models. 
It is unclear whether this behavior affects significantly the evolution of the instability. 
Note also that the ion tracer show a different thickness in the two different shear layer (central panel), due to the cumulative effect of the (fluid) velocity shear and (particle) ion gyromotion \cite{Nakamuraetal2010POP}. 
%-------------------------------------------------

%================================
\begin{figure}[b!]
\begin{center}
 \includegraphics[width=\linewidth]{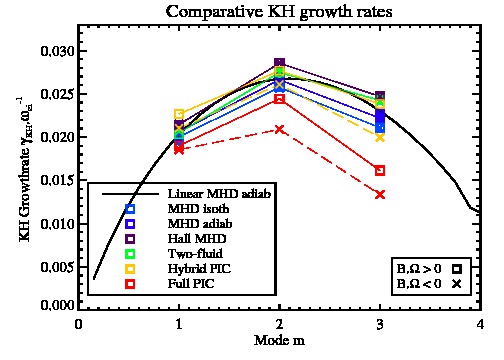}
\end{center} 
\caption{Comparative Kelvin-Helmholtz growth rates for different models using the same initial setup. Squares and full lines (resp. crosses and dash lines) are used to represent the growth rate computed at the shear layer where $\mathbf{B} \cdot \mathbf{\Omega} > 0$ (resp. $\mathbf{B} \cdot \mathbf{\Omega} < 0$).}
\label{fig:comparativeGrowthRates}
\end{figure} 
%================================

%--------------------------------
\subsection{Linear growth rate} \label{section:linearphase}
%--------------------------------
The different codes first run with the previously described initial conditions until the end of the linear phase of the instability. 

In Fig.~\ref{fig:comparativeGrowthRates} we show the growth rates for the first three unstable modes, computed using the different models, 
with the growth rate of the KHI (over-plotted with a black continuous line) calculated from the linear MHD theory with an adiabatic closure, 
using the linear MHD LEDAFLOW code \cite{Nijboeretal1997CPC}. In the figure the cross and square symbols refer to the relative orientation 
of the vorticity with respect to the magnetic field. 
The fastest growing mode in this setup is the mode $m=2$, which corresponds to a wavenumber $k_{_{FGM}} d_i = 1/30$. 
The different models show a good agreement at large wavelength (mode m=1) and differ at smaller wavelengths (modes m=2 and m=3). 
This is a first signature of the expected disagreement between different plasma descriptions, when kinetic and ion inertial scales are reached. 

Hereafter, we will refer to the MPI-AMRVAC code (section~\ref{codes_amrvac}) to model ideal MHD, and to the Stagger code (section~\ref{codes_stagger}) to model resistive MHD. 
The growth rates obtained using the two MHD codes with an adiabatic closure are the same and identical to the growth rate resulting from the linear MHD 
(adiabatic) theory.
In a compressible fluid, part of the available energy also goes to compression. This is why the growth rate is slightly smaller when the closure is isothermal, since the compressibility is larger. 

The MHD, Hall MHD and two-fluid models show very similar results during the linear stage of the instability, 
before the nonlinear dynamic of the vortices create gradients at scales of the order of the ion inertial length. 
The growth rates obtained from the Hall-MHD and two-fluid simulations are however slightly larger than those obtained from the MHD simulations. 
The Hall effect modifies the effective magnetic tension since the magnetic field lines are now frozen-in in the electron fluid and no more in the ion fluid. 
Thus, the stabilizing effect of the magnetic tension is reduced in the presence of the Hall term, explaining why the KHI growth rate is slightly larger when 
using the Hall-MHD or the two-fluid model. 

Squares (resp. crosses) are used in Fig.~\ref{fig:comparativeGrowthRates} to represent the growth rate computed at the shear layer where $\mathbf{B} \cdot \mathbf{\Omega} > 0$ (resp. $\mathbf{B} \cdot \mathbf{\Omega} < 0$). The relative orientation of the velocity field vorticity $\Omega$ compared to the background magnetic field $B_z$ does not modify the KH growth rate in the MHD, 
Hall-MHD and two-fluid regimes, as shown in Fig.~\ref{fig:comparativeGrowthRates}. 
Indeed, the fluid models used in this study do not contain intrinsic information on the particle gyration. 
On the contrary, models that include ion kinetic effects (Hybrid and full PIC codes represented by yellow and red lines in Fig.~\ref{fig:comparativeGrowthRates}) 
show that the relative orientation of $\Omega$ and $B_z$ modifies the linear dynamics of the KHI. 
Note that different fluid models that take into account Finite Larmor Radius (FLR) corrections in a fluid description \cite{PassotSulemHunana2012POP} 
are expected to exhibit the same asymmetry in the KHI growth rate, observed here with hybrid and full kinetic models, as shown in previous studies with FLR-MHD models\cite{Huba1996GRL}. 
In particular, we observe that both hybrid and full PIC calculations show the same qualitative feature: 
the growth rate for $\mathbf{B} \cdot \mathbf{\Omega} < 0$ is smaller than that for $\mathbf{B} \cdot \mathbf{\Omega} > 0$, 
as observed also in previous PIC simulations \cite{WilberWinglee1995JGR,Nakamuraetal2010POP}. 
This result is in contrast with previous analytic calculations including FLR effects in an extended MHD description \cite{Nagano1978JPlPh,Nagano1979PSS}, 
where the growth rate was found to be larger for $\mathbf{B} \cdot \mathbf{\Omega} < 0$ than for $\mathbf{B} \cdot \mathbf{\Omega} > 0$.
It is particularly striking that all kinetic simulations exhibit the contrary. This is most probably due to the lack of initial kinetic equilibrium 
that implies a fast readjustment of the kinetic plasma which modifies the effective setup before the instability starts (see section.~\ref{section:kineticrelaxation}). 
The discrepancies observed in the growth rate computed from the hybrid and full kinetic descriptions are most probably due to the treatment of electrons as a massless isotropic isotermal fluid in the first case and kinetically in the second case. 
In this setup, electrons and ions have the same initial temperature so that the total pressure is initially equally divided between them. In this case, the electron compressibility is also expected to play a key role. 
The high level of noise in PIC codes, which implies a greater uncertainty on the calculation of the KHI growth rates, could also account partly for the growth rate 
discrepancies using the kinetic models.

Note that the ion inertia should also affect the growth rates depending on the relative orientation of $\Omega$ and $\mathbf{B}$\cite{FujimotoTerasawa1991JGR}. 
However, at least in the range of parameters used in this study, the Hall term do not give any measurable difference in the growth rate between the $\mathbf{B} \cdot \mathbf{\Omega} > 0$ and $\mathbf{B} \cdot \mathbf{\Omega} < 0$ cases. 
On the other hand, the hybrid and full kinetic models show different growth rates for different orientation of 
the vorticity and the magnetic field. Therefore, since in these models both the Hall and the FLR effects are included, 
we conclude that the difference observed in the growth rates for the two orientation is dominated by the FLR effects. 

Interestingly, including ion inertial effects (spatial scale $d_i$) in the model increases the growth rates, as seen when comparing (adiabatic) MHD growth rates to (adiabatic) Hall-MHD and two-fluid growth rates. 
On the contrary, including ion kinetic effects by means of a kinetic code (in particular the ion gyroradius $\rho_i$) decreases the growth rates whatever the relative orientation of $\Omega$ and $B$. 

The observed difference between fluid and kinetic growth rates could be a consequence of different effects: (i) the absence of initial kinetic equilibrium, which could induce significant modifications in the shear layer, (ii) a difference in the plasma compressibility between the fluid and kinetic models, (iii) the influence of finite ion Larmor radius effects. 

%--------------------------------
 \begin{figure*}%[!h]
 \centering \includegraphics[width=0.9\linewidth]{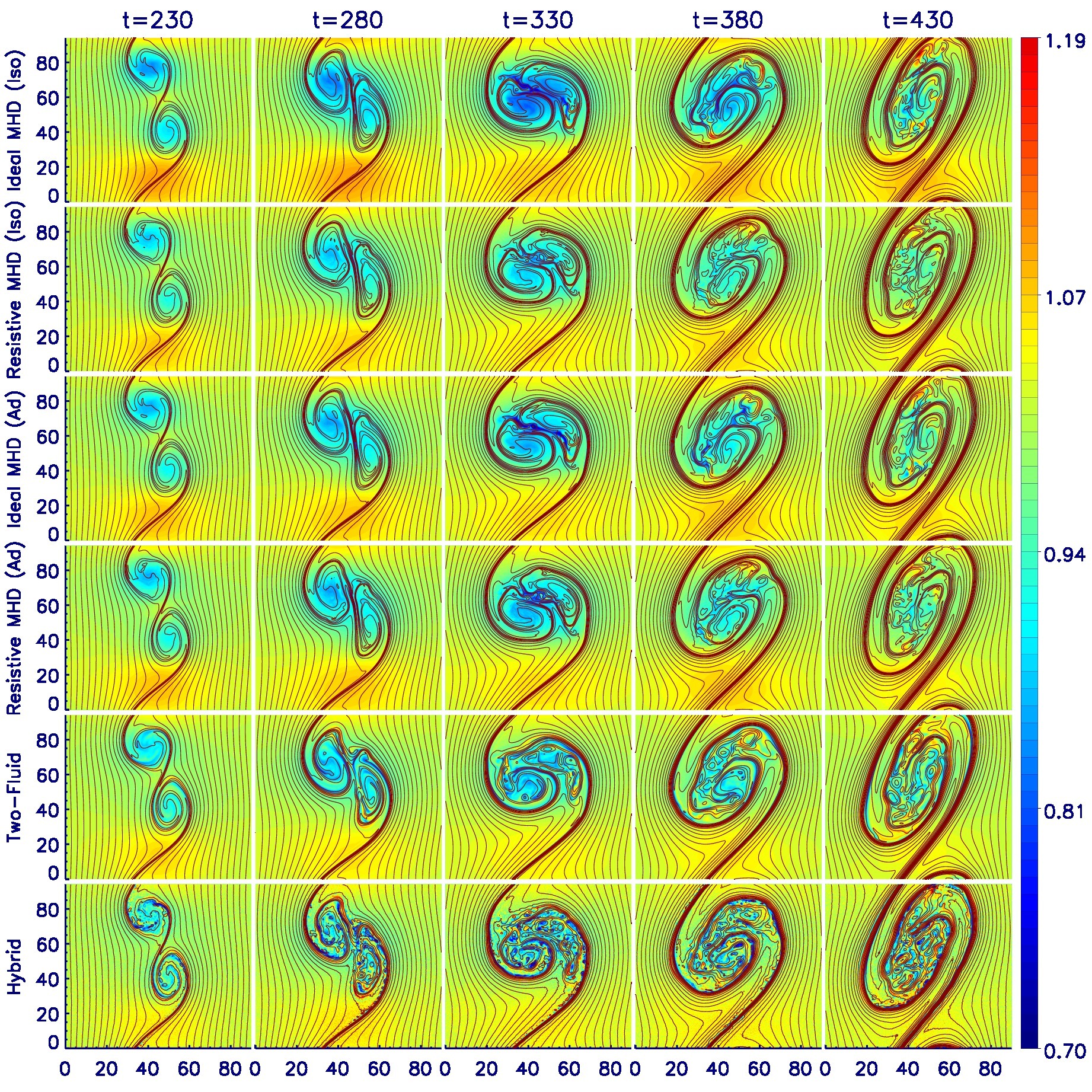}
 \caption{Evolution of the density $n$ with superimposed magnetic field lines for the shear layer centered on $x \simeq 45\ d_i$, 
 characterized by $\mathbf{B} \cdot \mathbf{\Omega} > 0$. The time evolves from left to right: $t \omega_{ci}=230$, $280$, $330$, $380$ and $430$ 
 for each respective column. From top to bottom, each line shows a single model: isothermal ideal MHD, isothermal resistive MHD, adiabatic ideal MHD, 
 adiabatic resistive MHD, two-fluid and hybrid models. The axes, shown at the bottom and on the right for sake of clarity, 
 represent the $xy$-positions expressed in ion inertial length. }
  \label{fig:All_Den_BWpositive}
 \end{figure*}
%--------------------------------

%--------------------------------
 \begin{figure*}%[!h]
 \centering \includegraphics[width=0.9\linewidth]{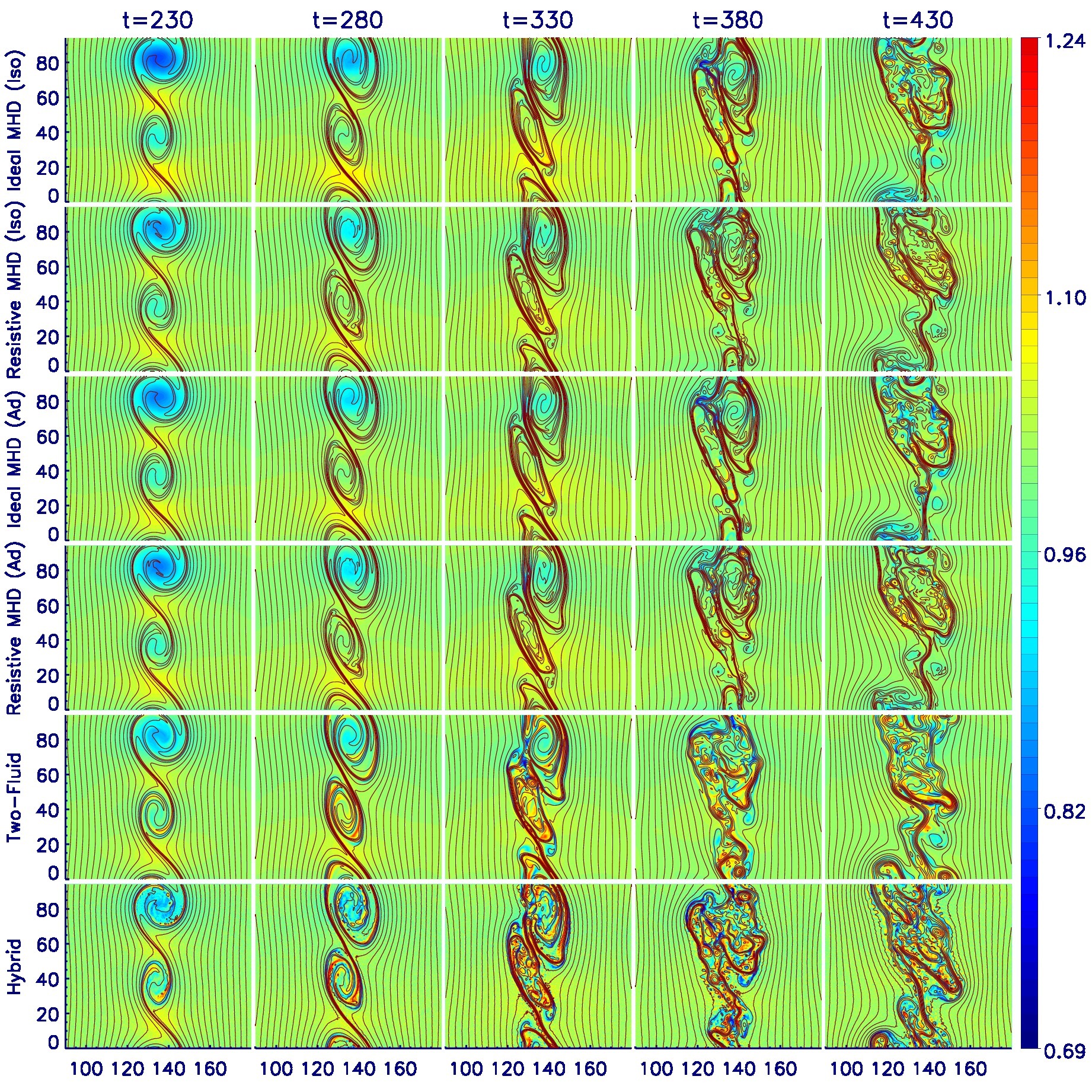}
\caption{Evolution of the density $n$ with superimposed magnetic field lines for the shear layer centered on $x \simeq 135\ d_i$, characterized by $\mathbf{B} \cdot \mathbf{\Omega} < 0$. The time evolves from left to right: $t \omega_{ci}=230$, $280$, $330$, $380$ and $430$ for each respective column. From top to bottom, each line shows a single model: isothermal ideal MHD, isothermal resistive MHD, adiabatic ideal MHD, adiabatic resistive MHD, two-fluid and hybrid models. The axes, shown at the bottom and on the right for sake of clarity, represent the $xy$-positions expressed in ion inertial length. }
  \label{fig:All_Den_BWnegative}
 \end{figure*}
%--------------------------------

%################################################
%--------------------------------
\subsection{Nonlinear evolution} \label{section:nonlinearphase}
%--------------------------------

%--- Introduction - phases - models - no full PIC
In this section, we consider the following models: isothermal ideal MHD, isothermal resistive MHD, adiabatic ideal MHD, adiabatic resistive MHD, two-fluid and hybrid models. The full kinetic (PIC) model is not used in this part because of the huge computational time required to model electrons kinetically. A full kinetic simulation of the KHI is a computational challenge that will be tackled in future works. To properly compare the nonlinear evolution of the KHI obtained from different models, we consider \emph{identical} initial velocity perturbations in the different runs by imposing the same first random phases $\Phi_m$, $m=1,\dots,6$, however different from one layer to the other. The initial amplitude of the perturbations is now set to $\epsilon=0.05$. 

%----------------------------------------------------------------
 \subsubsection*{Comparative nonlinear evolution at large scales}
%----------------------------------------------------------------

The nonlinear evolution of the KHI is shown in Figs.~\ref{fig:All_Den_BWpositive} and~\ref{fig:All_Den_BWnegative} (layers centered at $x \simeq 45\ d_i$ and $x \simeq 135\ d_i$, characterized by $\mathbf{B} \cdot \mathbf{\Omega} > 0$ and $\mathbf{B} \cdot \mathbf{\Omega} < 0$ respectively), through the evolution of the density and of the magnetic field lines. For sake of clarity, both layers are plotted separately on different pictures, Figs.~\ref{fig:All_Den_BWpositive} and~\ref{fig:All_Den_BWnegative} respectively; however we recall that the simulations are periodic in both directions. Each row corresponds to a single model, while time evolves from left to right. 

%--- Differences between models -- large scale -- Differences between layers -- large scale
At the beginning of the nonlinear phase ($t \omega_{ci}=230$, first column in Figs.~\ref{fig:All_Den_BWpositive} and~\ref{fig:All_Den_BWnegative}), two fully rolled-up vortices develop along each shear layer. 
The differences between the two shear layers are mainly due to the different phases $\phi_m$ used in the initial perturbations. 
At this early nonlinear stage, before smale scale processes develop, no appreciable discrepancies arise between the different models for a given shear layer. 

%--- 
The classical evolution of the hydrodynamic KHI, in the absence of magnetic field, is an inverse cascade characterized by the merging of the different vortices generated from the fastest linear growing mode, up to the formation of a single large vortex in the numerical box\cite{Baty2003PhPl}. 
In the magnetized case reported here, the onset of vortex pairing is shown at $t \omega_{ci}=280$ (second column in Fig.~\ref{fig:All_Den_BWpositive}), on the first shear layer, $x \simeq 45\ d_i$, with $\mathbf{B} \cdot \mathbf{\Omega} > 0$. The merging from two rolled-up vortices to a single one then develops at successive times $t \omega_{ci}=330$ and $380$ (third and fourth columns). 
On the contrary, in the case $\mathbf{B} \cdot \mathbf{\Omega} < 0$ (Fig.~\ref{fig:All_Den_BWnegative}), no vortex merging is observed. Instead, at $t \omega_{ci}=280$ (second column), the vortices remain aligned with the initial shear layer direction at $x \simeq 135\ d_i$. At later times, the onset of vortex pairing is observed at $t \omega_{ci}=330$ (third column) but small scale processes disrupt the vortices before the pairing is completed, leading to the formation of a mixing layer instead of a large single vortex. 

To summarize the global non linear dynamics observed in the two different layers, we observe that in the first layer the final state is given by a large scale, coherent single vortex, while in the second layer, the final state is given by a turbulent layer. Both cases are well captured at large scale by all the considered models, from ideal MHD to hybrid kinetic.

In Figs.~\ref{fig:All_Den_BWpositive} and~\ref{fig:All_Den_BWnegative}, the $y$-direction is represented in order to give a better feeling on the matching (or mismatchings) between the successive models. We recall that the numerical box is periodic. 
Note that the different models perfectly connect one another at the early stage of the KH nonlinear evolution (first column) and connect well at large scales during the nonlinear evolution, showing the agreement between all models at large scales. 
Note also that some discrepancies arise at small scale in the late stage of the KH nonlinear evolution (last column in particular), as can be seen when looking at the `boundaries' between two successive models. 

%--------------------------------
 \begin{figure*}
 \centering \includegraphics[width=0.85\linewidth]{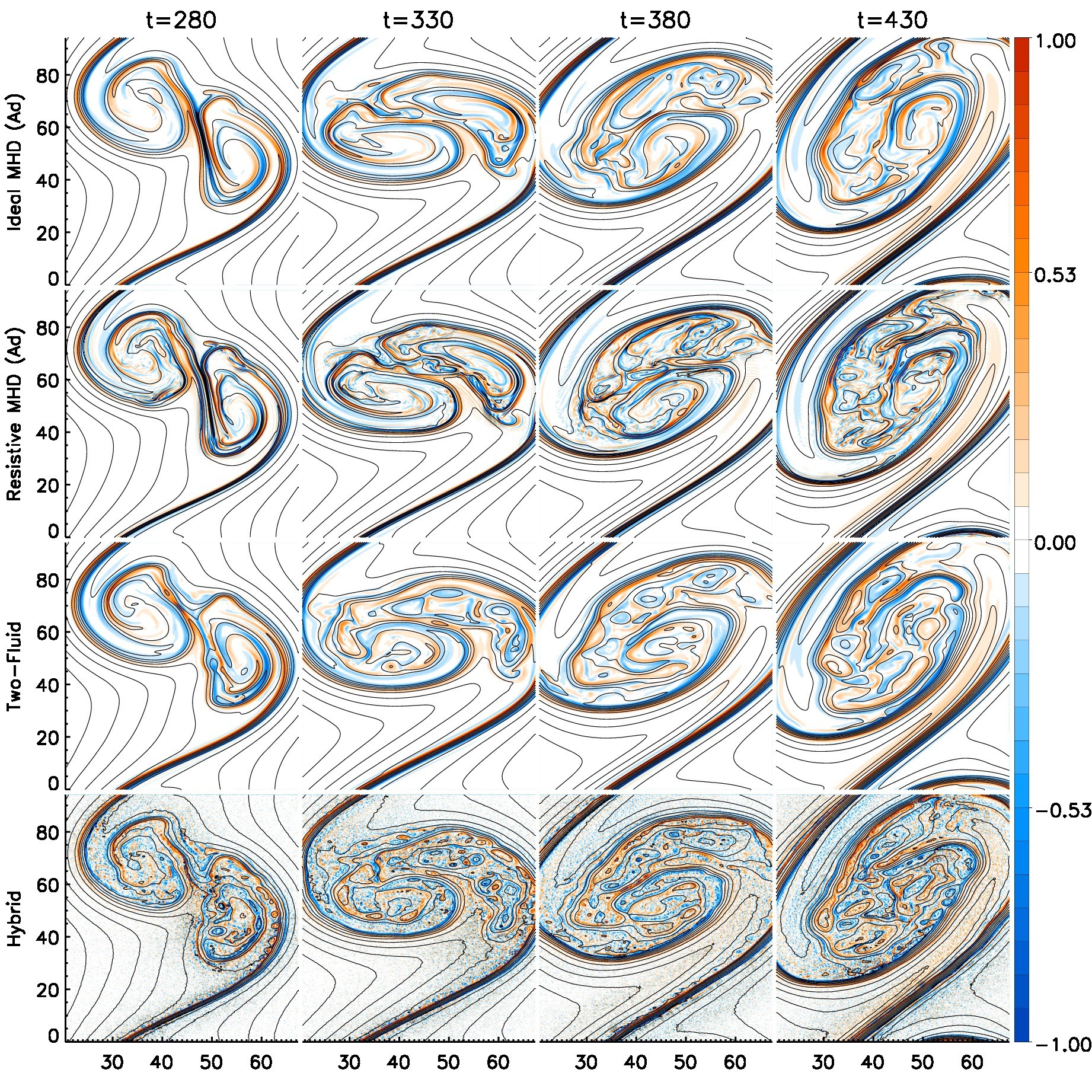}
  \caption{Evolution of the out-of-plane current $J_z$ with superimposed magnetic field lines for the shear layer centered on $x \simeq 45\ d_i$, characterized by $\mathbf{B} \cdot \mathbf{\Omega} > 0$. The time evolves from left to right: $t \omega_{ci}=280$, $330$, $380$ and $430$ for each respective column. From top to bottom, each line shows a single model: adiabatic ideal MHD, adiabatic resistive MHD, two-fluid and hybrid models. The axes, shown at the bottom and on the right for sake of clarity, represent the $xy$-positions expressed in ion inertial length. The color dynamics is saturated to better identify the current structures. }
  \label{fig:All_Jz_BWpositive}
 \end{figure*}
%--------------------------------

%--------------------------------
 \begin{figure*}
 \centering \includegraphics[width=0.85\linewidth]{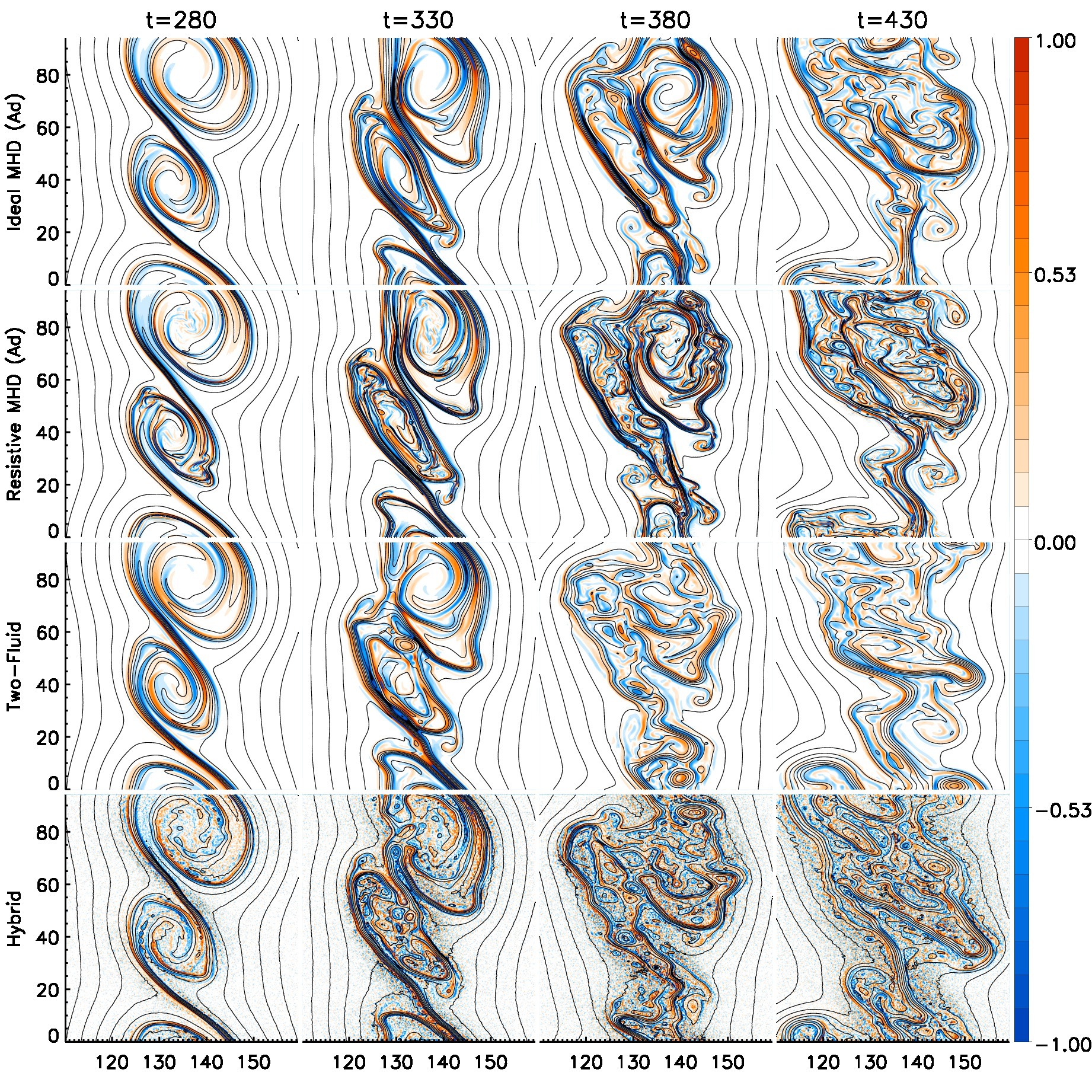}
 \caption{Evolution of the out-of-plane current $J_z$ with superimposed magnetic field lines for the shear layer centered on $x \simeq 135\ d_i$, characterized by $\mathbf{B} \cdot \mathbf{\Omega} < 0$. The time evolves from left to right: $t \omega_{ci}=280$, $330$, $380$ and $430$ for each respective column. From top to bottom, each line shows a single model: adiabatic ideal MHD, adiabatic resistive MHD, two-fluid and hybrid models. The axes, shown at the bottom and on the right for sake of clarity, represent the xy-positions expressed in ion inertial length. The color dynamics is saturated to better identify the current structures. }
  \label{fig:All_Jz_BWnegative}
 \end{figure*}
%--------------------------------

%----------------------------------------------------------------------------------------------------
\subsubsection*{Comparative nonlinear evolution at small scales: the essential role of magnetic reconnection}
%----------------------------------------------------------------

%--- Differences between models -- small scales
At scales much smaller than the hydrodynamic scale $L_{_{HD}}$, the magnetic field is stretched and compressed by the plasma motion, forming current sheets between and inside the vortices, at scales of the order or smaller than the ion inertial length $d_i$ (Figs. \ref{fig:All_Jz_BWpositive} and \ref{fig:All_Jz_BWnegative}). In the ideal and resistive MHD simulations, strong current sheets are formed but magnetic reconnection does not start at early stage. On the contrary, the same current sheets modeled by the two-fluid and the hybrid models become tearing unstable and develop magnetic reconnection. 
Such a difference at small scales is due to the Hall effect, included in the two-fluid and the hybrid models, but absent from the ideal and resistive MHD models. The Hall effect allows for magnetic reconnection to develop at a faster rate in the two-fluid and the hybrid models, while the MHD models rely on the explicit (resistive MHD code) or the numerical (ideal MHD code) resistivity for reconnection to occur \cite{Birnetal2001JGR}. 

In the nonlinear phase of the magnetized KHI, two regions are typically subject to reconnection: (i) the compressed region that separates two merging vortices and (ii) the sheared magnetic regions inside the vortices themselves. 
The Hall effect enables the tearing instability to develop faster in both regions: (i) between the two pairing vortices, as seen by the X-points and magnetic islands formed at $x d_i^{-1} = 50$ at $t \omega_{ci}=280$ (first column) in Fig.~\ref{fig:All_Jz_BWpositive} and at $x d_i^{-1} \simeq 135$ at $t \omega_{ci}=330$ (second column) in Fig.~\ref{fig:All_Jz_BWnegative} for the two-fluid and hybrid models (bottom two panels) in contrast to the surviving magnetic shear in the two MHD models (top two panels) at the same times; (ii) inside the single vortices, as seen by the chains of magnetic islands (e.g. at $x d_i^{-1} = 60$, $y d_i^{-1} = 60$ at $t \omega_{ci}=330$ (second column) in Fig.~\ref{fig:All_Jz_BWpositive} and at $x d_i^{-1} \simeq 125$ at $t \omega_{ci}=330$ (second column) in Fig.~\ref{fig:All_Jz_BWnegative}. 
Note that vortex-induced reconnection still occurs in resistive MHD in both regions but at later times, $x d_i^{-1} = 60$, $y d_i^{-1} = 40$ at $t \omega_{ci}=330$ and $y d_i^{-1} \simeq 70$ at $t \omega_{ci}=380$ in Fig.~\ref{fig:All_Jz_BWnegative}. 
In the hybrid simulation, the reconnection process seems to be triggered earlier with respect to the two-fluid simulations; as seen by the early formation of chains of magnetic islands inside the vortices in Fig.~\ref{fig:All_Jz_BWpositive} (bottom left panel, $x d_i^{-1} = 65$). This may be due to the intrinsic higher level of noise in particle simulations that seeds the tearing instability at a higher initial level, possibly through non-modal transient amplification, and makes it saturate much earlier.

%#################################################
\section{Discussions and conclusions} \label{section:conclusions}
%#################################################

%--------------------------------
In this paper we have reported the first comparison of the magnetized Kelvin-Helmholtz instability, during its linear and nonlinear evolution. We have used several different plasma fluid and kinetic models: an isothermal/adiabatic ideal/resistive MHD, Hall-MHD, a two-fluid, a hybrid kinetic and a full kinetic model. \\

%---------------------------------------------------------
In the linear stage of the KHI, the MHD models do not care about the relative orientation of the vorticity with respect to the magnetic field direction, while the fluid simulations that include ion inertia (Hall-MHD, two-fluid) are insensitive to it; on the other hand, 
kinetic simulations (Hybrid and Full PIC) clearly exhibit different growth rates depending on this relative orientation, showing that FLR effects dominate ion inertia, within the parameter range used in this study. \\

%--------------------------------------------------------
With the parameters used in this benchmark, the fastest growing mode m=2 builds up two rolled-up vortices in the nonlinear stage of the instability, that eventually merge into a single vortex or are disrupted before merging because of secondary magnetic reconnection. 
At the end of the simulations, even if differences are present between the two sides of the layer, a common feature arises: although the KH vortices are large-scale MHD structures at the beginning of the nonlinear phase, their nonlinear evolution eventually leads to the formation of smaller length scales, both inside the large $m=1$ merged MHD vortex (left side) or in the mixing layer (right side). Even if small scale processes are strongly dependent on the choice of the model, interestingly, the global large-scale picture is well captured by all the considered models. 
Note however that the degree of plasma mixing and the development of turbulence may strongly vary from one model to another at later times due to the nature of the small scale process at play. \\

%--------------------------------------------------------
The final stage of the instability is significantly different between the two shear layers, and all models capture the same final state, at least using a coarse graining view. 
Since the differences between the evolution of the two layers are also captured by the MHD model, we conjecture that such differences are not driven by kinetic or inertial effects, in particular not by the relative orientation of the vorticity with respect to the magnetic field. Previous works in the context of the MHD modeling of the KHI have shown that the nonlinear interactions differ dramatically, as influenced through the imposed phase differences between linear modes \cite{Baty2003PhPl}. In the setup described here, the initial phases of the first perturbed modes are fixed among the various models but are different for both shear layers. 
The origin of the two different final states in the two shear layer could rely upon the fact that the linear perturbations on both sides differ, as influenced by the initial phases, and can hence nonlinearly evolve differently. 

%--------------------------------------------------------
The difference arising for a given model between the two different shear layers (see the differences between 
Figs.~\ref{fig:All_Den_BWpositive} and~\ref{fig:All_Den_BWnegative} considering a same row) is much more important than 
the difference arising between different models on the same shear layer (see the last columns of 
Figs.~\ref{fig:All_Den_BWpositive} and~\ref{fig:All_Den_BWnegative}). This indicates that the large 
scale, fluid structure is much more influenced by the choice of the initial phases than by the small scale processes at play. 
Moreover, the fluid models are shown to capture the large scale dynamics as well as the kinetic model, at least in the regime of parameters used in this study. 
This indicates that the feedback of small (inertial and kinetic) scales to the large scale dynamics does not appear as a dominant process even in the nonlinear phase of the magnetized shear flow. 
This study emphasizes the importance of the fluid behavior in the nonlinear evolution of magnetized shear flows, which determines the large scale structures and the saturation of the vortices. This result should strongly encourage the development of fluid codes to study the nonlinear dynamics of magnetized plasmas at large, fluid scales. 
Note however that the complementary use of kinetic models is necessary to carry out the validation of the fluid approach at large scales.   \\

%--------------------------------
We must also stress the important consequences of the different closures used in the different codes. 
The MHD and two-fluid codes both use a standard adiabatic closure, using a polytropic law with polytropic index $\gamma = 5/3$ (on electrons also in the two-fluid model). The electrons are treated as an isothermal fluid in the hybrid PIC code. 
On the contrary, no explicit evolution law is imposed on the ions (hybrid PIC, full PIC) and electrons (full PIC), 
so that the behavior of effective ion (and electron for the full PIC code) compressibility remains a priori unknown 
and may change in space and time. We recall that the compressibility plays a key role in the development of the KHI. 
In particular, it modifies the growth rate of the instability. Finding a clever closure on the fluid codes, 
in accordance with the compressibility found in the kinetic simulations, will enable to properly compare the growth rate, 
in a first step, and the nonlinear dynamics of the KHI, in a second step. \\  

%--------------------------------
In order to accurately describe the cross-scale, nonlinear evolution of collisionless plasmas, the coupling between different plasma models is showing recently an increasing interest from the plasma community \cite{SwiffArticle}. 
In this context, comparisons of numerical simulations from different models, as the one described in this paper, 
represent a necessary step before coupling codes from different plasma models. 
In order to properly couple different codes, it is a necessity to make sure that the different physical ingredients, introduced by the different models, describe the common features of interest at the large scales. 
We have shown in this work that the large, fluid scales are little disturb by the small, kinetic scales, since fluid and kinetic simulations capture the same behavior at large scales. Such a result is likely to strongly encourage the development of multi-scale code coupling for colisionless plasmas.  \\

%--------------------------------
We stress here that the numerical modeling of plasma dynamics is a fundamental problem of major interest in present plasma physics research. 
Therefore, this study is of broad interest and is not limited to the KHI itself and related nonlinear dynamics. 
We thus underline two important aspects of the fluid and kinetic modelling of the nonlinear evolution of magnetized plasma. 
From a fluid point of view, the question of the fluid closure that plays an important role in the dynamics is shown not to be trivial in a magnetized plasma. 
From a kinetic point of view, we have illustrated the necessity to find a correct initial setup that takes into account the kinetic effects at play, such as finite Larmor radius effects. These fundamental problems need to be addressed in future works. \\
%################################################

\noindent
{\bf Acknowledgment} The research leading to these results has received funding from the European Commission's Seventh Framework Programme (FP7/2007Ð2013) under the grant agreement SWIFF (project n 263340, www.swiff.eu). This work was supported by the Italian Super-computing Center CINECA under the ISCRA initiative. This work was supported by the HPC-EUROPA2 project (project number: 228398) with the support of the European Commission - Capacities Area - Research Infrastructures. UNIPI acknowledge the access to the HPC resources of CINECA made available within the Distributed European Computing Initiative by the PRACE-2IP, receiving funding from the European Community's Seventh Framework Programme (FP7/2007-2013) under grant agreement n$^\circ$ RI-283493 and project number 2012071282. Work at ASI and IAP, ASCR was also supported by projects RVO: 67985815 and RVO: 68378289. This work was supported by NASA award NNX11A1164G. UCPH acknowledges computer resources provided by the Danish Center for Scientific Computing (DCSC), now part of the Danish e-Infrastructure Collaboration (DeIC).

% \bibliographystyle{aipnum4-1}
% \bibliography{/home/phenri/Documents/COMPTE_RENDUS/mabiblio}


\end{document}